\documentclass[fleqn,usenatbib]{mnras}

\usepackage{mathptmx}
\usepackage{txfonts}

\usepackage[T1]{fontenc}
\usepackage{ae,aecompl}

\usepackage{graphicx}	
\usepackage{amssymb}	

\usepackage{pdflscape}	
\usepackage{newtxtext}

\newcommand{\gaia}{\emph{Gaia}}

\title{\gaia\ pulsars and where to find them}

\author[J. Antoniadis]{John Antoniadis$^{1,2,3}$\thanks{E-mail: janton@mpifr.de} 
\\
$^{1}$Max-Planck-Instutut f\"{u}r Radioastronomie, Auf dem H\"{u}gel 69, 53121, Bonn, DE\\
$^{2}$Argelander Institut f\"{u}r Astronomie, Auf dem H\"{u}gel 71, 53121, Bonn, DE\\
$^{3}$Institute of Astrophysics, FORTH, Dept. of Physics, University Campus, GR-71003 Heraklion, Greece  
}

\pubyear{2020}

\begin{document}
\label{firstpage}
\pagerange{\pageref{firstpage}--\pageref{lastpage}}
\maketitle

\begin{abstract}
While the majority of massive stars have a stellar companion, most pulsars appear to be isolated. 
Taken at face value, this  suggests that most massive binaries break apart due to  strong natal kicks received in supernova explosions. 
However, the observed binary fraction can still be subject to strong selection effects, as monitoring of newly-discovered pulsars is rarely carried out for long enough to conclusively rule out multiplicity.  
Here, we use the second \gaia\ Data Release (DR2) to search for  companions to 1534  
rotation-powered pulsars with  positions known to better than 0\farcs5. 
We find 22 matches to known pulsars, including 1 not reported elsewhere, 
and 8 new possible companions to young pulsars.
We examine the photometric and kinematic properties of these systems and 
provide empirical relations for identifying \gaia\ sources with potential millisecond pulsar companions.  
Our results confirm that the observed multiplicity fraction is small. However, we 
show that the number of binaries below the sensitivity of \gaia\ 
and radio timing in our sample could still be significantly higher. We 
constrain the binary fraction of young pulsars to be $f_{\rm young}^{\rm 
true}\leq 5.3(8.3)\%$ under realistic(conservative) assumptions for the binary 
properties and current sensitivity thresholds. For massive stars ($\geq 10$\,M$_{\odot}$) in particular, we find  $f_{\rm OB}^{\rm 
true}\leq 3.7\%$ which sets a firm independent upper limit on the galactic neutron-star merger rate, $\leq 7.2\times 10^{-4}$\,yr$^{-1}$. 
Ongoing and future projects such 
as the CHIME/pulsar program, MeerTime, HIRAX and ultimately the SKA, will 
significantly improve these constraints in the  future.    
\end{abstract}


\section{Introduction} \label{sec:intro}
Multiplicity plays a critical role in the evolution of massive stars:  
interactions with a stellar companion can modify  a star's mass, composition and angular momentum, thereby  influencing the manner in which it dies, and the type of remnant it produces.  
Studies of  star-forming associations suggest that most young 
massive stars ($\sim 100\%$) have a 
gravitationally-bound companion, with the majority ($\sim 70$\%) being   close enough to eventually interact via mass exchange \citep{Sana:2012px}. Such interactions can lead to stellar mergers, X-ray binaries (XRBs), supernovae (SNe) and hyper-velocity stars. Systems that survive these processes and remain bound  ultimately become compact-object binaries and gravitational wave (GW) sources.

Probing how the multiplicity fraction evolves as a stellar population ages is crucial for understanding these astrophysical phenomena and the 
underlying physical processes, including natal SN kicks \citep{bailes1989,PortegiesZwart:1999cn,janka2013}, the physics of 
common envelope (CE) evolution \citep{ivanova2013}, and the birth vs remnant mass relation \citep{antoniadis2016b,Tauris:2017omb}. 

Theoretical models and predictions should confront to  a number of observational benchmarks, such as the  distribution of pulsar velocities \citep{hobbs2005}, the number of XRBs and their relative distances from star-forming regions \citep{Walter:2015zta,Tauris:2017omb}, the birth rates and properties of double neutron star systems \citep{kalogera2004,Tauris:2017omb,ferdman2020}, and the observed GW merger rates \citep{abadie2010,abbott2017}.   
One important independent constraint that remains underutilized is the  
multiplicity fraction among young, non-recycled pulsars \citep{bailes1989}. The frequency of binary systems that survive SN explosions is a sensitive 
probe of the SN kick magnitude distribution, which remains one of the major theoretical uncertainties. There are only a handful 
of known binaries hosting young  radio pulsars, implying  that most 
SNe result in disruptions. However, the observed multiplicity fraction is 
 likely subject to significant selection effects as most new pulsar 
discoveries are typically  monitored only for up to a few months. Owing to the 
presence of timing noise and the strong co-variances with the intrinsic spin-down, such observations are not particularly sensitive to orbital 
acceleration caused by low-mass companions and/or wide binary pairs. 
Indeed, even some of the few known binaries have been discovered only by chance \citep{lyne2015,Kaplan:2016ymq,Bassa:2016fiy}. 

In this work we revisit the constraints on  pulsar multiplicity  using the second \gaia\ data release \citep[\texttt{DR2;}][]{gaiacollaboration2016,Lindegren:2018cgr} which, for the first time, provides information on the distances and kinematic properties of nearly two billion stars in the Galaxy and Magellanic Clouds. In Section~\ref{sec:2} we describe our sample and methodology, while in Section~\ref{sec:3} we present our results. In Section~\ref{sec:4} we examine the influence of various selection effects and current limitations, and derive a constraint on the multiplicity fraction of young pulsars. We conclude with a discussion in Section~\ref{sec:5}.   

\section{Methods}\label{sec:2} 

\subsection{Sample}\label{sec:2.1}
In what follows, we distinguish between different classes of rotation-powered pulsars based on their inferred minimum dipole magnetic field strength, $(B/{\rm G})\equiv 3.2\times 10^{19} (P\dot{P} s^{-1})^{1/2}$. More specifically, sources with $(B/{\rm G})\leq 10^{9}$ are considered fully-recycled millisecond pulsars (MSPs), the ones with  $10^{9}<(B/{\rm G})\leq 10^{10}$ are mildly-recycled pulsars (henceforth mild MSPs), while those with $(B/{\rm G})>10^{10}$ are referred to as young, non-recycled pulsars. 

\begin{figure}

 \includegraphics[width=\columnwidth]{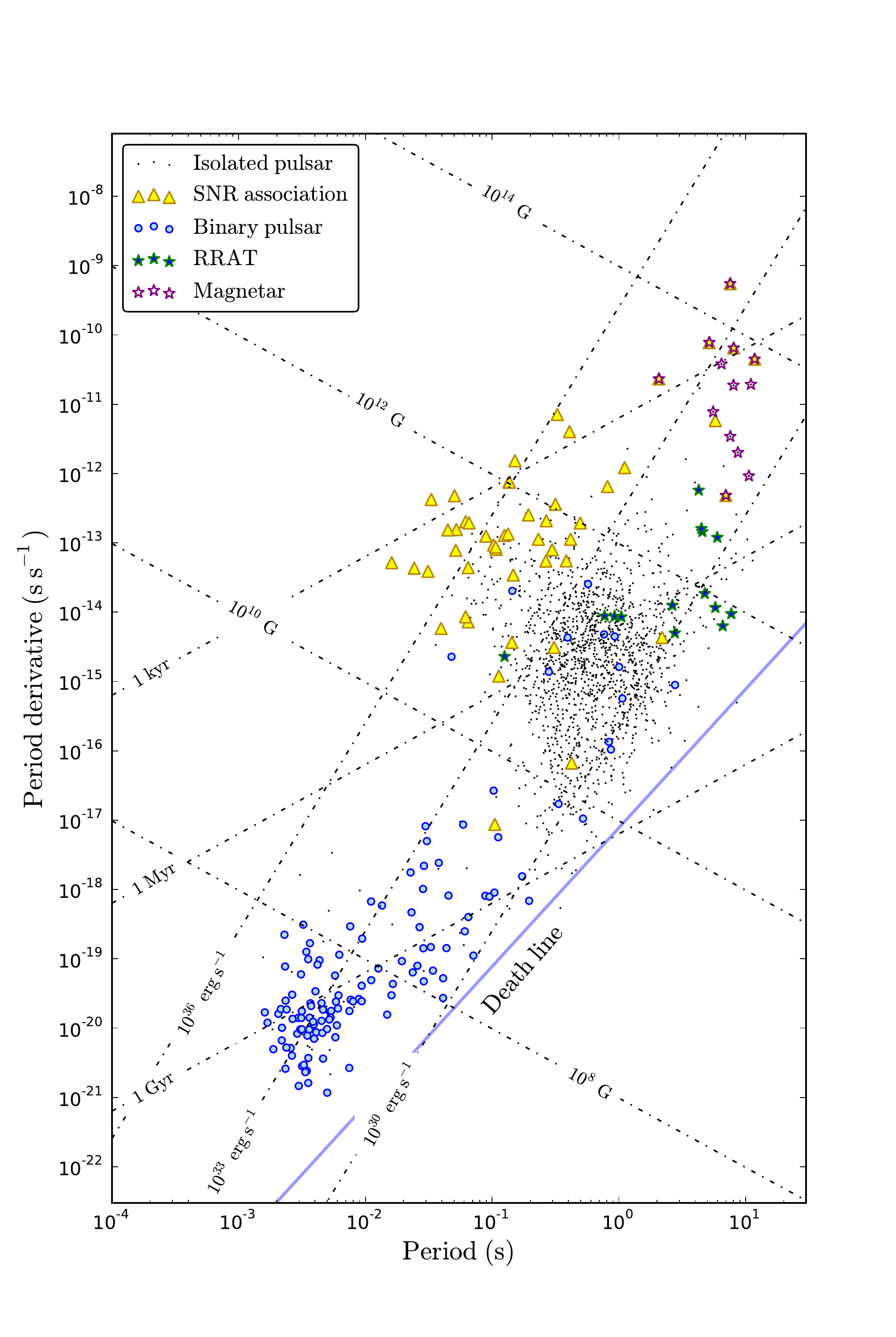}
 \caption{P--$\dot{P}$ diagram of the pulsars contained in our sample.}
 \label{fig1}
\end{figure}

Our main goal is to identify possible companions to known rotation-powered pulsars. There are currently 2492  pulsars outside of globular clusters listed in the  ATNF pulsar 
catalogue\footnote{\url{http://www.atnf.csiro.au/research/pulsar/psrcat} accessed on Oct\,20\,2020} \citep[\texttt{PSRCat v1.63;}][]{Manchester:2004bp}. Since we are primarily interested in 
finding close astrometric pairs, we only considered those with 
positions known  to better than 0\farcs5. This leaves 1534 objects (see 
Figure\,\ref{fig1}), a sample considerably larger than those used in previous 
searches for \gaia\ counterparts (65 MSPs in \citealt{mingarelli2018}, 155 
binaries in \citealt{jennings2018} and 57 non-recycled pulsars in \citealt{Igoshev:2019cwq}). Of these, 107 are MSPs, 39 are mild MSPs and  1388  are young  pulsars. The sample contains 145 known binaries (86 MSPs, 32 mild MSPs and 27   
young pulsars) and 11 extra-galactic sources. 

\subsection{Identification of Candidate Astrometric Pairs}\label{sec2.2}
We  searched for possible pulsar companions among  \gaia\ DR2 objects \citep{Lindegren:2018cgr} located within 20\arcsec\ of each pulsar.
This search radius is equivalent to the angular distance that would be covered by an object at 500\,pc,  moving with a transverse 
velocity of  $v_{\rm tr}=1200$\,km\,s$^{-1}$ for 40\,yr.   
For each \gaia\ entry, we  used the available astrometry to propagate the celestial coordinates  and 
their uncertainties to the reference epoch of the  pulsar ephemeris (\texttt{PosEpoch} in \texttt{PSRCat}). 
This choice allowed us to utilize the entire sample instead of only the subset of pulsars with  timing  proper motions.   For the few sources for which only a two-parameter astrometric solution is available, we  used directly their \gaia\ epoch position  (J2015.5).  
Sources outside the 2$\sigma$ positional error circle of the pulsar were dismissed as unrelated.  
For each possible astrometric pair,  we inferred a probability of chance coincidence using the average number density of stars within 1\arcmin\ of the pulsar \citep{antoniadis2011}. 

Given that many pulsar and \gaia\ DR2 positions considered here are known to $\ll 0\farcs5$ precision, their uncertainties are likely  dominated by systematics. 
One important source of potential errors is  rotations between the \gaia\ reference 
frame --- which is tied to the  International Celestial Reference Frame \citep[ICRS, see][]{mignard2018} --- 
and the various incarnations of the Solar System ephemeris utilized in pulsar timing. Co-variances 
between astrometric and other parameters in the pulsar timing solution, red noise 
in the \gaia\ astrometry and timing noise may play a role as well. Here, we did not attempt to correct explicitly for these effects. Instead, we  compensated for them by adding $0\farcs25$ in 
quadrature to the inferred position error. This value was adopted    solely based on empirical  
considerations. More specifically, it corresponds to $\sim 1/2$ of the second largest separation between a pulsar and its known optical counterpart (PSR\,J0045$-$7319; see Table~\ref{tab:1} and Section~\ref{sec:3.2}). 
This choice also made our search  sensitive to nearby  astrometrically-resolved binaries that would have been missed otherwise ($0\farcs25$ is equivalent to a separation of $125$\,AU at 500\,pc, see Section~\ref{sec:3.1}). Increasing the systematic uncertainty beyond this value --- for instance, to include PSR\,J2129$-$0429 for which the inferred separation is $\sim 1\farcs4$ --- resulted in a very large number of false positives.

\subsection{Distances, Reddening and Kinematics}
It is well known that obtaining distance estimates by directly inverting the parallax measurement can lead to significant bias in the presence of noise.  
Following previous work \citep{jennings2018} and recommendations \citep[][]{bailer-jones2015,astraatmadja2016,bailer-jones2018,luri2018}, we inferred probability density functions (PDFs) for the distance to sources of interest using: 
\begin{equation}\label{eq:1}
    P(d|\hat{\pi}) \propto \frac{1}{2L^3}e^{-d/L}d^2 e^{(-1/d - \hat{\pi})^2 /2\sigma^2_{\hat{\pi}}},
\end{equation}
Here, we assume that the parallax measurement, $\hat{\pi}$ is normally distributed about the true parallax, $1/d$, with a dispersion $\sigma_{\hat{\pi}}$. The $e^{-d/L}d^2L^{-3}$ term accounts for a Lutz-Kerker bias \citep{lutz1973} with an exponentially-decreasing stellar density (that is nearly constant for $d\ll L$). The former prior combined with the condition $d>0$ has the advantage of making the distance PDF 
well-behaved and normalisable so that statistical moments can be defined.  For our purposes, distances are not the primary quantity of interest 
and thus $L$ was simply taken to be direction-independent and was set equal to 1.5\,kpc  \citep{bailer-jones2018}. This should be adequate  for the 
majority of sources with small Galactic latitudes, but may overestimate the distances to sources with imprecise parallaxes located outside the 
Galactic plane.  For more detailed models, see \cite{jennings2018,bailer-jones2015} and references therein. 

When applicable, we also made use of other available measurements that contain distance information. More specifically, for certain pulsar/\gaia\ cross-matches we utilized timing and VLBI parallaxes (inverted using  Eq.\,\ref{eq:1}), and DM-based distance estimates derived using the \texttt{NE2001} model and assuming a 20\% uncertainty \citep[][]{deller2019,ding2020}. For extra-galactic sources we used the most recent distance estimates to the Large(Small) Magellanic Cloud, $49.97\pm1.13(62.1\pm1.9)$\,kpc \citep{pietrzynski2013,graczyk2013}. 
For non-certain associations, we simply required the distance be positive.
In what follows we report  median values as  point estimators together with  16\% and 68\% percentiles as uncertainties.  

The interstellar reddening along each line of sight was traced using the latest version of the 3D dust map of \cite{green2019} \citep[\texttt{Bayestar19}\footnote{\url{http://argonaut.skymaps.info/}}; see also][]{green2014,green2018}. Finally we derived estimates for the reduced proper motion \citep{luyten1922} using, 
\begin{equation}\label{eq:2}
    H_{\rm g} = m_{\rm g} + 5\log_{10}{\frac{\mu}{{\rm mas\,yr^{-1}}}}-10+A_{\rm g}  
\end{equation}
where $m_{\rm g}$ and $A_{\rm g}(d)$ are the apparent magnitude and  extinction in the \gaia\ band, and  $\mu = \sqrt{\mu_{\alpha}^2 + \mu_{\delta}^2}$ is the magnitude of the proper motion. This quantity is analogous to the absolute magnitude but sensitive to the source's transverse velocity, $H_{\rm g}\equiv M_{\rm g} + 5\log_{10}(
v_{\rm tr} /4.74057\,{\rm km}\,{\rm s}^{-1})$. This makes it a useful quantity  for distinguishing between different kinematic populations, even in the absence of a precise parallax measurement. 

\section{Results}\label{sec:3}
Our search identified 32 close astrometric pairs with a probability of chance coincidence smaller than 10\%. These sources include 
 22 known binaries, 8 new candidate counterparts, the Crab pulsar (PSR\,B0531+21) and the pulsar wind nebula of PSR\,B0540$-$69 in the LMC.  For the majority of these sources, the inferred probability of chance coincidence is $1-3\%$, implying that the number of false positives in this sample is  $\mathcal{O}(3)$ or smaller.
 
 The properties these \gaia\ sources are listed in Table\,\ref{tab:1}, while  Figure\,\ref{fig2} shows their location on color--magnitude (henceforth HR) and color--H$_{\rm g}$ (henceforth RPM) diagrams. 
 
As can be seen, for most confirmed pulsars, the angular separation $\hat{\theta}$ at \texttt{PosEpoch} is much smaller than the position error,  which is dominated by the systematics. Two notable exceptions are PSR\,J0045$-$7319, for which the timing solution has not been updated for over 25 years \citep{kaspi1996a}, and PSR\,J0437$-$4715, a nearby MSP. The reason for the large discrepancy in the latter case is not clear, but it might be related to the faintness of the optical counterpart and the proximity of the system to Earth.
 
 In the rest of this section we discuss the properties of confirmed matches and candidates in more detail. 
 
 \subsection{Millisecond Pulsars}\label{sec:3.1}
We found  18 matches to known MSPs including PSR\,J0348+0432 which, even though does not satisfy our formal MSP definition,  is most likely a fully-recycled pulsar \citep{Antoniadis:2013pzd}.  With the exception of PSRs\,B1953+29, B1957+20 and J1435$-$6100, these \gaia\ counterparts have also been identified by \cite{jennings2018}, \cite{mingarelli2018} and \cite{Igoshev:2019cwq}. 

\begin{figure*}
\begin{tabular}{cc}
 \includegraphics[width=\columnwidth]{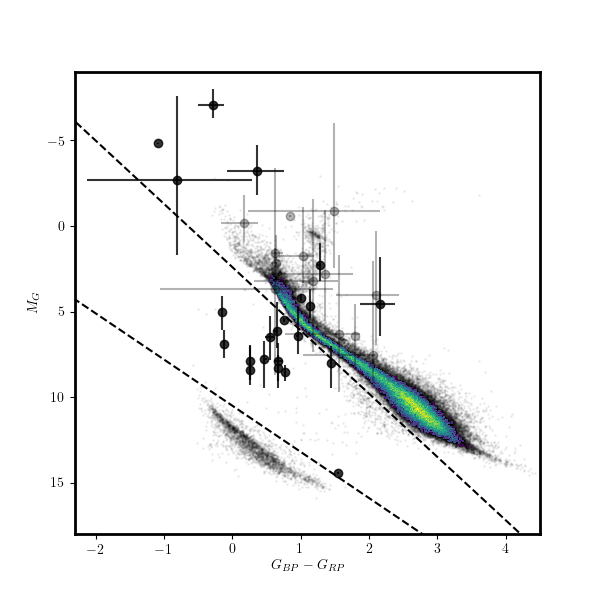} & \includegraphics[width=\columnwidth]{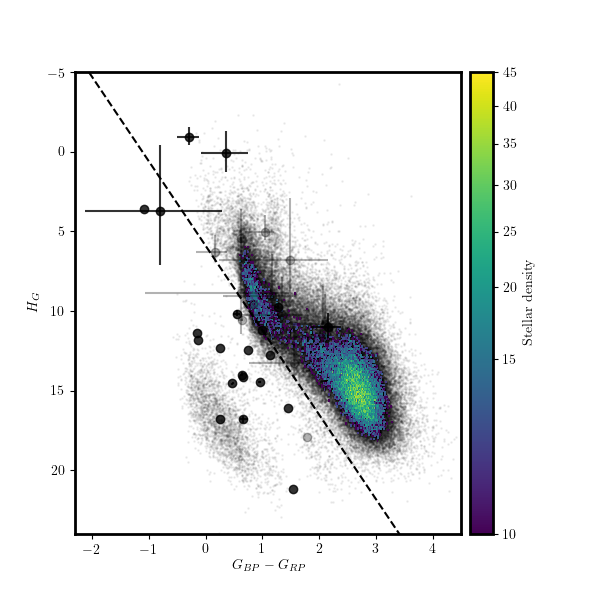} 
\end{tabular}
 \caption{Positions of known and candidate pulsar companions and with \gaia\ counterparts on the observational HR (left) and RPM diagrams (right).  Scatter points and the overlaid histogram shows field stars with precise parallax measurements within 200\,pc from the Sun. The positions of known pulsars and candidate associations are shown  in black and gray respectively. The dashed lines show the empirical cuts of  Eq.~\ref{eq:3} (see text for details). }
 \label{fig2}
\end{figure*}

The majority of these sources are eclipsing MSPs with stellar- or Jupiter- mass companions (often referred to as ``redbacks'' and ``black widows'' respectively). As can be seen in Figure~\ref{fig2} these sources occupy sparsely populated regions on the HR and RPM diagrams: they are both bluer and significantly brighter compared to MS stars of comparable masses. This supports the idea that they either have a high helium content (i.e. they are stripped), or that their atmospheres are highly irradiated. On the RPM diagram, MSP companions are even more separated from MS stars. On average, they have large  proper-motion moduli, suggestive of high  velocities that are likely a fossil signature of the SN explosion.  
We find that most eclipsing MSPs, as well as  three MSPs with low-mass white dwarf (WD) companions, satisfy the following conditions: 
\begin{equation}\label{eq:3}
\begin{array}{rcl}
M_{\rm G}  & \leq   & 2.7(G_{\rm BP}-G_{\rm RP}) + 10.5,  \\
M_{\rm G}  & >  & 3.7(G_{\rm BP}-G_{\rm RP}) + 2.4, \\ 
H_{\rm g}  & >  & 5.3(G_{\rm BP}-G_{\rm RP}) + 5.9, \\
\end{array}
\end{equation}
 where $G_{\rm BP}$ and $G_{\rm RP}$ are the mean magnitudes in the \gaia\ BP and RP photometric bands. 
These empirical criteria are  efficient in filtering out 99.8\% of the field stars shown in Figure~\ref{fig2}. They are also  robust against selecting high-velocity MS  stars with similar $H_{\rm g}$ values. However, some overlap is expected with early-type sub-dwarfs and extremely-low mass white dwarfs and their precursors \citep{pelisoli2019b,pelisoli2019c}. 
Utilizing color and H$_{\rm g}$ information only (i.e. the last two relations) results in significant overlap with WDs and higher contamination from halo stars (at the $\sim 1-2\%$ level),  but  has the  benefit of being applicable to sources for which a reliable  parallax measurement is not available.  
This can be compensated by additional criteria (e.g. $H_{\rm g}   \leq   6.9(G_{\rm BP}-G_{\rm RP}) + 13.0$) to filter out WDs, or 
multi-wavelength information (e.g. the presence of X-ray, $\gamma-$ray or radio counterparts). We provide some practical examples for identifying pulsar candidates in a python notebook accompanying this 
paper\footnote{10.5281/zenodo.4075042}. 

\subsubsection*{Individual sources and comparison to previous work}
 \paragraph*{PSR\,J1024$-$0719} 
 This MSP was long thought to be 
 isolated. However \cite{Bassa:2016fiy} and \cite{Kaplan:2016ymq} identified a $\sim 0.4$\,M$_{\odot}$ MS 
 K-dwarf as a potential companion 
 in a very long period (2-20 kyr) 
 orbit. Our search identified \gaia\,\texttt{3775277872387310208} as the pulsar companion \citep[see also][]{jennings2018} 
 with a true angular separation of $\theta=0\farcs076(3)$ (without 
 accounting for systematic uncertainties which are likely not dominant in this case). At 
 the inferred distance of $3.4^{+4.6}_{-2.0}$\,kpc, this 
 implies a projected  separation of $260^{+852}_{-195}$\,AU, or an orbital period between $\sim 0.75$ and 12\,kyr. The uncertainty is clearly dominated by the  distance estimate and thus it will  improve significantly with further timing observations and the next \gaia\ data release. 
This source clearly demonstrates the complementary of  optical astrometry and radio timing for identifying binary pulsars. 

\paragraph*{PSR\,B1957+20} This pulsar is the prototypical ``black widow'' system, a 2-ms MSP with a Jupiter-mass companion in a 9\,h orbit. \cite{jennings2018} discuss a counterpart $\sim 0\farcs7$ away from the timing position but they conclude that this source is unrelated. Here, we identify a faint source within 0\farcs1 of the pulsar (\gaia\,\texttt{1823773960079216896}). We are confident that this is the  counterpart to PSR\,B1957+20 based on available optical photometry  \citep[e.g.][]{reynolds2007} and the good match between the \gaia\ and timing proper motion measurements. Combining the negative parallax measurement with DM-based distance estimate, yields $d=1.9^{+1.0}_{-0.9}$\,kpc, consistent with the constraints based on  spectroscopy \citep{vankerkwijk2011}.  

\paragraph*{PSR\,B1953+29} This system is the second MSP discovered after PSR\,B1937+21 and the first one in a binary \citep{boriakoff1983}. It has a low-mass companion \citep[$\lesssim 0.18$\,M$_{\odot}$;][]{gonzalez2011} in a 117\,d orbit. 
Source \gaia\,\texttt{\,2028584968839606784} was identified as a possible counterpart based on 
its  angular separation  of $0\farcs1$ from the pulsar position, and the low stellar density in the field which implies a small probability for chance alignment ($\leq 3$\%). However, the error budget is dominated by the systematic uncertainty which is likely overestimated, given that the timing position is being updated regularly \citep{arzoumanian2018,arzoumanian2020}. 
The same source has  been investigated as the possible companion to PSR\,B1953+29 by \cite{boriakoff1996} and \cite{mignani2014}. It has a red featureless spectrum suggestive of an evolved star.
Indeed, the inferred colors and absolute luminosity place the source above the MS. 
However, the \gaia\ parallax measurement implies a distance of $2.3^{+2.9}_{-1.0}$\,kpc which is $\sim 2$ times smaller than the DM-based estimate. In addition, the proper motion does not agree well with that of the pulsar. 
We therefore conclude that the \gaia\ source is probably unrelated to the 
pulsar and that the true companion of the latter is a low-mass WD, as suggested by the radio timing solution. 

\paragraph*{PSR\,J1435$-$6100} This is a 9\,ms pulsar with a massive, $\sim 1$\,M$_{\odot}$ companion in a 32\,h orbit. 
This particular combination of fast spin, short orbital period and high companion mass is challenging to explain. The compactness of the orbit suggests that the system is the survivor of a CE episode. 
To reach the observed spin frequency, the pulsar must have accreted $\mathcal{O}(0.1$\,M$_{\odot}$) of material from the donor star \citep{podsiadlowski2002,tauris2012a}. This happened either during the CE event or after, via Case BB mass transfer from the stripped, post-CE companion. In either case, the short evolutionary timescales suggest that the system experienced a phase of supercritical accretion in which a very high accretion flow that led to efficient neutrino cooling \citep{tauris2011a,tauris2012a}.

 We identified a 18.9$^{m}$ source (\gaia\ \texttt{5878387705005976832})  0\farcs3 away from the pulsar timing position. The object has a probability of true association $\sim 95\%$, and a trigonometric parallax that is consistent with the pulsar DM distance estimate. However, the inferred absolute magnitude of $M_{\rm G}\simeq -2.7^{\rm m}$ means that this cannot be the companion responsible for recycling the pulsar. Deep photometric observations obtained by \cite{jacoby2006} failed to detect any fainter counterpart closer to the pulsar position, which is consistent with the expected WD nature of the companion. Based on that, we conclude that this is either a chance coincidence or, less likely, the system is part of a wide triple. The latter possibility can be tested if the pulsar proper motion is measured.

\paragraph*{PSRs J1311$-$3430, J1628$-$3205 and J2129$-$0429}
\gaia\ counterparts to these systems were identified by \cite{jennings2018}, but 
 did not show up in our search. The position of PSR\,J1628$-$3205 is not well known and hence, it was not included in our final sample. For PSR\,J1311$-$3430,  the nearest source was filtered out due to its large position error of $2\farcs9$. Similarly, for PSR\,J2129$-$0429, the inferred angular separation of the nearest source (1\farcs3 for \gaia\ \texttt{2672030065446134656}) was larger than our formal $2\sigma$  threshold. However, for all these cases the association is confirmed by phase-resolved photometry and spectroscopy. We include these sources in Table~\ref{tab:1} for completeness.

\paragraph*{PSRs J1843$-$1113, J1732$-$5049 and J1949+3106}
Possible associations to these pulsars were discussed by \cite{mingarelli2018}. For the nearest source to PSR\,J1843$-$1113 (Gaia\,\texttt{4106823440438736384}), we infer a separation of 1\farcs44. Long-term timing of this pulsar \citep{desvignes2016} strongly suggests that it is isolated. Hence, we conclude that the \gaia\ source is most likely unrelated. 

For PSRs J1732$-$5049 and J1949+3106 which are known binaries, we again infer large separations to the proposed counterparts (2\farcs26 and 2\farcs82 for \gaia\,\texttt{5946288492263176704} and \texttt{2033684267592307584} respectively). In addition, the inferred colors, luminosities and proper motions are inconsistent with the white-dwarf nature of the pulsar companions and the pulsar timing constraints. We therefore conclude that these sources are also unrelated.

\subsection{Young and mildly-recycled pulsars}\label{sec:3.2}
Of the 1440 pulsars with $B>10^9$\,G in our sample, 11 were found to have a close astrometric pair. Of these, only three are known binaries: 
PSR\,J0045$-$7319  is an LMC pulsar with a B1V companion in a 51\,day orbit \citep{kaspi1996a,kaspi1996b}. PSR\,B1259$-$63 is in a highly-eccentric 3.4-yr orbit with a rapidly-rotating OB star \citep{chernyakova2014}. The system is well known for its 40-day long radio eclipses near periastron. 
Finally, PSR J2032+4127 is a pulsar in the Cygnus OB2 association. It is in a $\sim$30-yr orbit around a Be star. It's binary nature was revealed only after long-term timing \citep{lyne2015}.  

In addition to the aforementioned sources, our sample contained a further 59 young and mildly-recycled pulsars in known binaries. Upon investigation we find that all their companions were  too faint to be detected by \gaia\ \citep{vankerkwijk2005,mignani2014}. 
We identified 8 new  close astrometric pairs to which we discuss below. 
\subsubsection*{Candidates}

\paragraph*{PSR J0534$-$6703} This source is a slow pulsar in the LMC \citep{manchester2006}. We found a bright star within $\sim 0\farcs5$ of the timing position. Its propoerties are consistent with an early-type MS star at the LMC. The positional offset  is similar to the one between PSR\,J0045$-$7319 and its optical counterpart. Given that the ATNF timing solution is over 15 years old, this could be attributed to systematics. Interestingly, the pulsar has an unusually high spin period derivative, which might  be caused by orbital acceleration. 
We conclude that a true association between the pulsar and the \gaia\ source is plausible and cannot be ruled out based on existing information. We therefore  recommend further radio timing observations.  

\paragraph*{PSR\,J1624$-$4411}
The nearest counterpart to PSR\,J1624$-$4411 has a positional offset of 0\farcs3 from the timing position \citep{lorimer2006}. The \gaia\ source is consistent with an MS star at the inferred distance of $2.7^{+4.6}_{-1.8}$\,kpc. The latter also agrees well with the DM-based estimate of $\sim 3.3$\,kpc and hence a true association is possible. 

\paragraph*{PSR\,J1638$-$4608}
\gaia\ \texttt{5992089022760118400} has a separation of only 0\farcs14 from the timing position of this pulsar. Unfortunately, only a 2-parameter astrometric solution is available and hence the true separation might be larger. At the DM-distance of the pulsar, the source would have an absolute magnitude of $M_{\rm G}=0.3^{+1.5}_{-1.6}$.  
Interestingly, \cite{kerr2016} report a 450-day periodic modulation in the timing residuals. If interpreted as a R{\o}emer, the observed amplitude would suggest a planetary companion of a few Earth masses. Such an object would be much fainter than the \gaia\ source. Under these considerations, we conclude that this is either a chance alignment or that the \gaia\ source is in a very wide orbit. The next \gaia\ data releases will hopefully result in more stringent constraints on this system. 

\paragraph*{PSR\,J1838$-$0549}
The nearest \gaia\ entry has an angular separation of $\sim0\farcs5$ from the pulsar. Its distance is somewhat smaller but still consistent with the pulsar's DM distance. The inferred distance implies that source is a giant star off the MS. \cite{parthasarathy2019} recently presented timing data spanning 4 years that show no signs of periodicity. Based on the available information, we conclude that the \gaia\ source is either in a wide orbit around the pulsar or a foreground star.

\paragraph*{PSR\,J1852+0040}
PSR\,J1852+0040 is a radio-quiet pulsar associated with the SN remnant Kes 79 \citep{halpern2010}. The \gaia\ source identified in our search has a distance that is only marginally consistent with that of the pulsar. Hence, we conclude that this is either a chance coincidence or that the source is the ejected companion of the pulsar progenitor. 

\paragraph*{PSR\,J1903$-$0258}
This pulsar is located 0\farcs4 away from \gaia\,\texttt{4261581076409458304} \citep{lorimer2006}. The measured color and luminosity suggest that the latter is an intermediate mass  MS star. 
Given that the timing solution has not been updated for over 15 years, the two objects may be physically associated.

\paragraph*{PSR\,J1958+2846} 
This is a radio-quiet pulsar that has been timed using the Fermi-LAT instrument. The $\gamma-$ray timing solution \citep{kerr2016} derived using $\sim1.5$-yr of data shows no sign of binarity. However, given the low accuracy of the time-of-arrival measurements and the small span of the data, an association with \gaia\,\texttt{2030000280820200960}, located $\sim 0\farcs4$  from the formal position is still possible. 

\paragraph*{PSR\,J2027+4557}
This pulsar was found in a radio survey of  Cygnus OB associations \citep{janssen2009}. 
The \gaia\ source that showed up in our search is consistent with an early-type MS star at the inferred distance of $1.96^{+0.25}_{-0.20}$\,kpc. The latter is smaller than the DM distance estimate.   
The pulsar bears similarities to PSR\,J2032+4127, located inside the Cygnus OB2 region. The latter has a DM of 115\,pc\,cm$^{-3}$, i.e. $\sim 2$ times smaller. Hence, a local un-modelled contribution to the DM would be needed, which is not impossible given the complexity of the region.  

\section{The multiplicity fraction among young pulsars}\label{sec:4}
The results presented in Section~\ref{sec:3} confirm that the 
\emph{observed} multiplicity fraction (henceforth $f_{\rm young}^{\rm obs}$) 
among young pulsars is small: of the 1377 galactic pulsars\footnote{ (here we do not consider extragalactic pulsars, since some of the processes impacting $f_{\rm young}^{\rm true}$ depend strongly on  metallicity)} with 
$B>10^{10}$\,G in our sample, only 22(16)  are in binaries 
with(without) considering the candidates described in the previous 
section, thus $f_{\rm young}^{\rm obs}=1.16-1.59\%$.  The latter is 
considerably smaller than the multiplicity fraction among MSPs and 
mild-MSPs ($f_{\rm MSP}^{\rm obs}\simeq79\%$ and $f_{\rm mild}^{\rm obs} \simeq85\%$ respectively). 

In interpreting these results, one has to consider two strong  
selection effects: 1) Radio timing is not sensitive to 
binaries with orbital periods considerably longer than the observational 
timespan and R{\o}emer delays much smaller than the amplitude 
of the timing noise, and b) \gaia\ cannot detect sources fainter than 
$m_{\rm g}\gtrsim 20.5^{\rm m}$. 
To examine the impact of these  biases and place a constraint on the 
true multiplicity fraction among young pulsars, $f_{\rm young}^{\rm true}$,  we constructed a toy model to simulate various scenarios for 
the orbital and optical properties of possible companions stars and 
estimate their detectability at radio and optical wavelengths. 

 In these simulations, we split the 1377 young Galactic pulsars in our sample in three categories: the first contained all known binaries 
 ($P_{\rm b}\ge 0$ in \texttt{PSRCat}), the second  the seven Galactic candidates of Table~\ref{tab:1} and the third, the 
 remaining seemingly isolated pulsars.  We then used a Monte-Carlo sampling scheme with $10^5$ iterations to estimate the number of binaries in each group and infer the probability distribution for $f_{\rm young}^{\rm true}$. 
 All pulsars belonging to the first group were obviously always counted as binaries. For the second group,   we 
 decided on whether the candidates of Table~\ref{tab:1} are true binary companions using their inferred 
probabilities of chance coincidence.
For the third group, we  assigned a companion with a 
separation and mass drawn from various astrophysically-motivated 
distributions (see below). The brightness was inferred using a 
mass-luminosity relation for MS stars, the extinction curves of 
\cite{green2019} (see Section\,\ref{sec:3}), and a distance normally 
distributed about the DM distance with a fractional uncertainty of 20\%. 
The multiplicity fraction for each sample population was inferred by
counting all binaries in the first two categories and adding those systems from the final category that had orbital periods and apparent 
magnitudes above a certain detectability threshold, $P_{\rm b}\geq P_{\rm b}^{\rm thres}$ and $m_{\rm g}\geq m_{\rm g}^{\rm thres}$.

\begin{figure*}
 \includegraphics[width=2.2\columnwidth]{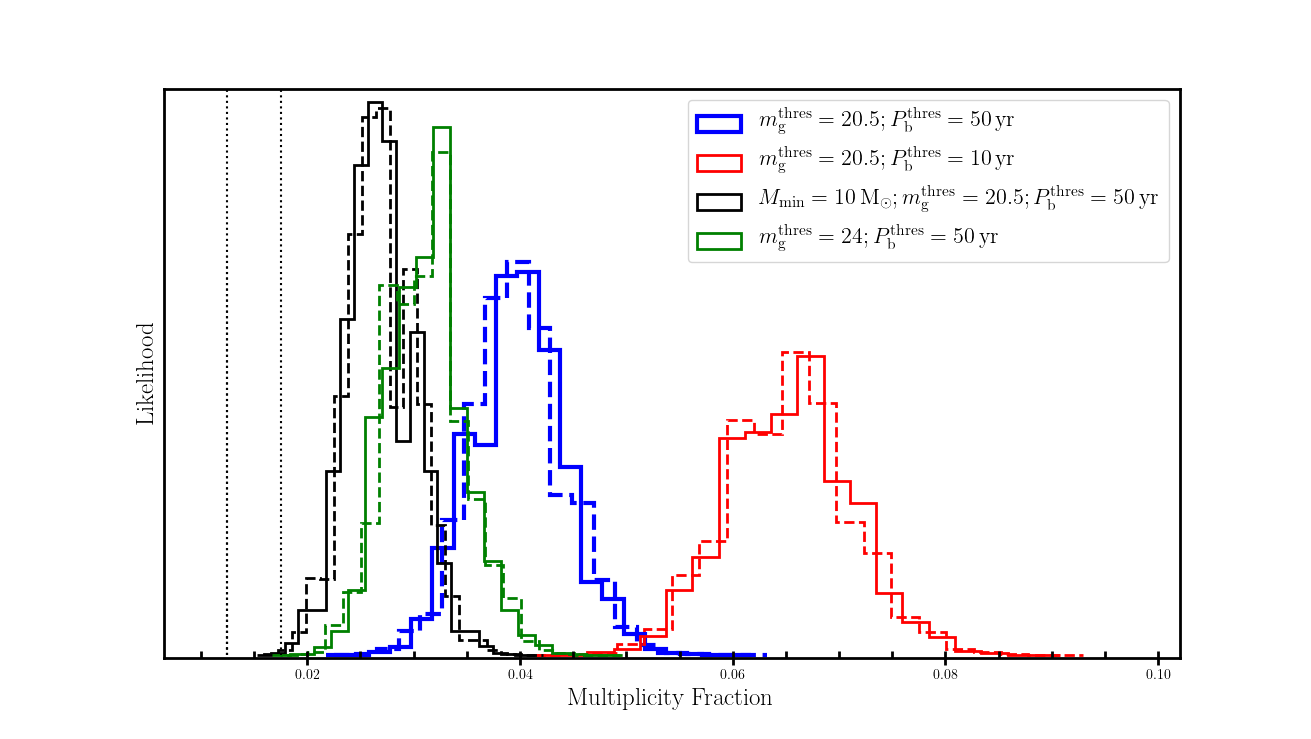}   
 \caption{The probability density function for $f_{\rm young}^{\rm true}$ given different detectability thresholds and scenarios discussed in the text (see Sections~\ref{sec:4} and \ref{sec:5}). Dashed lines show the results of simulations in which the possibility of free-free absorbtion of pulsar emission from the stellar wind is taken into account.}
 \label{fig3}
\end{figure*}

Figure~\ref{fig3} shows the results for various simulated scenarios.
For our fiducial model (shown in blue), companion masses were inferred by combining a Salpeter mass function in the 10\ldots50\,M$_{\odot}$ range for the pulsar progenitors, with the best-fit mass ratio distribution of \cite{Sana:2012px}. Orbital periods ranging from 1\,d  to 1000\,yr, were also distributed according to the results of the latter work, while for the \gaia\ and radio timing detectability thresholds we assumed $P_{\rm b}^{\rm thres}=50$\,yr and  $m_{\rm g}^{\rm thres}=20.5^{\rm m}$. This scenario --- which we consider to be the most realistic, given all underlying uncertainties --- places an upper limit of  $b_{\rm young}^{\rm true}\leq 5.3\%$ at the 99.8\% confidence level (CL). 
Radio timing appears to be the most sensitive probe of binarity ($f_{\rm young}^{\rm true}\leq 8.3\%$ at 99.8\% CL for $P_{\rm b}^{\rm thres}=10$\,yr), while massive companions would be the easiest to detect with \gaia\ ($f_{\rm young}^{\rm OB}\leq 3.6\%$ at 99.8\% CL for early-type companions brighter than $M_{g}= -2.6^{\rm M}$). Overall, we find that the binary fraction could still be $\sim 3$ to 8 times higher than the observed one, given present limitations. 

An additional selection effect against pulsars in binaries that is unaccounted 
for in the aforementioned models is  the stellar wind of the companion which may 
be opaque to radio emission. The high mass-loss rates expected for upper-MS stars could create a dense environment leading to free-free absorption. For pulsars in our sample, this means that certain combinations of orbits and companion masses are ruled out (given that a pulsar \emph{is} observed). 
For a circular orbit (i.e. constant separation), the minimum period below which 21-cm radiation would be eclipsed by a 1000\,km\,s$^{-1}$ outflow with a temperature of $10^4$\,K \citep[][]{illarionov1975b} is:

\begin{equation}\label{eq4}
    P_{\rm b}^{\rm min} \simeq 0.28 \left( \frac{\dot{M}}{4.8\times 10^{-12} {\rm M_{\odot}\,yr^{-1}}}\right) \left(\frac{M_{\rm tot}}{{\rm M_{\odot}}}\right)^{-1}\,{\rm days}
\end{equation}
where $\dot{M}$ is the wind mass-loss rate  and M$_{\rm tot}$ is the total mass of the binary. 
In practice, we find that taking this effect into account in our toy model -- assuming standard mass-loss recipes for the MS \citep{Kippenhahn:1994wva} --- has a very small impact on the results described above (see dashed lines in 
Figure~\ref{fig3}). This is because free-free absorption only becomes relevant for large companion masses and short separations. For the models discussed here, most such systems are assumed to be within the sensitivity of exiting timing programs and are thus already ruled out. 
However, Eq.\,\ref{eq4}  also suggests that some rotation-powered pulsars 
with massive companions  are obscured at radio wavelengths and therefore, the  intrinsic 
binary fraction could be even larger. 
Some clues about the influence of 
this effect is obtained from the properties of known systems: Of the pulsars investigated here, only the 1237-day binary PSR\,B1259$-$63 is eclipsed over $\sim 3$\% of its orbit near periastron \citep{johnston1992,johnston1994,Johnston:2005eq}. Considering the shape of the mass function, this 
suggests that stellar winds should only affect a small fraction of pulsars when they approach their massive companions (either due to compact orbits or  high eccentricities). On the other hand, pulsars in sufficiently close 
orbits around massive stars would become wind-fed high-mass X-ray binaries
(HMXBs). Indeed, 
most known wind-fed HMXBs have orbital periods in the 3\ldots300\,d range 
and X-ray luminosities that roughly scale as 
$L_{\rm x}\propto P_{\rm b}^{-4/3}$, as expected for spherical accretion 
\citep[e.g.][and references therein]{Lutovinov:2013ga,Chaty:2014pua,Walter:2015zta,Tauris:2017omb}. 
However, these sources are accretion-powered and hence  not directly  
relevant to the population of pulsars considered here. 

To conclude, we believe that absorption is only relevant to a small 
fraction of pulsars orbiting massive companions with periods  of up to few years. To be hidden, these 
pulsars must approach their companions sufficiently close to intersect the optically-thick part of the wind, but not too close so as to become persistent HMXBs. 
In turn, this means that the constraints on $f_{\rm young}^{\rm true}$ inferred in this section should be a reliable proxy of the true frequency of pulsars in stellar binaries.

\section{Conclusions and Prospects}\label{sec:5}
We have performed a search for optical counterparts to 1534 rotation-powered pulsars.  By utilizing the \gaia\ DR2 catalogue and back-propagating its astrometric solution to the reference epoch of each pulsar's ephemeris (Section \ref{sec:2}), we were able to search for  close  astrometric pairs. 
We identified 20 previously known binary companions --- the majority of which are companions to MSPs --- and 8 tentative new matches to young pulsars. For these  candidates, we find that their \gaia\ parallaxes are broadly consistent with the pulsar distance estimates and recommend further timing observations to test if they are true binary companions. 

Most MSP companions in \gaia\ DR2 are early-type sub-dwarfs significantly bluer and more luminous than MS stars with similar masses. Their colours, luminosities and reduced proper motions sets them apart from the majority of stars on the HR and RPM diagrams. In Section~\ref{sec:3.1}, we inferred empirical color-magnitude-H$_{\rm g}$ criteria (Equation~\ref{eq:3}) that satisfy the observed 
properties of eclipsing and WD companions to MSPs and  filter out $\gtrsim 99.8\%$  of field stars. While some contamination is 
expected by sdA stars \citep{pelisoli2019a,pelisoli2019b,pelisoli2019c}, these selection criteria allow to single out \gaia\ sources that might be orbited by neutron stars. Deep targeted radio surveys of these candidates might be proven a viable path towards 
discovering new MSPs. Because of the information provided by \gaia, pulsars discovered in this way would be extremely valuable for a broad range of astrophysical inquiry \citep{Antoniadis:2012vy,mingarelli2018,arzoumanian2020}. For instance, MSPs with bright WD companions enable  precise NS mass measurements, which are important for constraining the mass function of compact stars and the equation-of-state of super-dense matter \citep[][]{Antoniadis:2013pzd,antoniadis2016b}. Similarly, eclipsing MSPs are key in understanding NS recycling and the end-states of low-mass XRBs \citep[e.g. ][and references therein]{vankerkwijk2011,tauris2015a,Strader:2018qbi}. 

In Section~\ref{sec:4}, we used a toy model to investigate the impact of observational selection effects on the detectability of pulsar companions. We constrained the true multiplicity fraction among young pulsars to be $\leq 5.3\%$ under the assumption that all binaries with orbital periods  shorter than 50 years and companions brighter than $20.5^{m}$ would have been detected by \gaia\ and ongoing timing programs. We also concluded that the number of pulsars with longer orbital periods that are obscured by the wind of their companions should be small.

Given that our toy model only considers stellar binaries, the number of pulsars with  planetary companions could still be considerably larger. Indeed, low-mass objects are considered as the possible cause of the low-amplitude periodic signals found in the timing residuals of several young pulsars.  \citep[e.g.][]{ray2011,parthasarathy2019}. With this caveat in mind, our constraints on $f_{\rm young}^{\rm true}$ should be reliable proxy for the number of young pulsars orbited by stars. While our analysis confirms that $f_{\rm young}^{\rm true}$ is low, we find that it could still be a factor 5--8 larger than previously thought. Hence, there could still be of order $\sim100$ hidden binaries in the known pulsar population. 
For massive companions in particular, we inferred a more stringent limit of $f_{\rm OB}^{\rm true}\leq 3.7\%$. Combined with the  Milky-Way core-collapse SN rate \citep[$r_{\rm SN}=0.01-0.02$\,yr$^{-1}$, see][and references therein]{diehl2006,Keane:2008jj}, this places a very conservative, but independent upper limit on the Galactic NS coalescence rate, 
$r_{\rm mergers}\lesssim f_{\rm OB}^{\rm true}\times r_{\rm SN}=(3.7-7.4)\times 10^{-4}$\,yr$^{-1}$ (with the caveat that some close binaries may be HMXBs, see Section~\ref{sec:4}). The former is  consistent with ({\bf but more conservative than}) the more direct constraints based on double neutron star systems \citep[][]{kalogera2004,Pol:2020tfz}, the observed number of HMXBs in the Galaxy \citep{Tauris:2017omb} and the advanced LIGO/Virgo constraint of $(3.2^{+4.9}_{-2.4})\times 10^{-5}$\,yr$^{-1}$  \citep{theligoscientificcollaboration2020}.  

There are several evolutionary pathways leading to the formation of isolated young pulsars. Theoretical models predict that  while $\sim$one third originate from single stars or binaries that merge during the MS \citep{PortegiesZwart:1999cn}, the majority must be  due to SNe in  massive binary systems. The relative number of pulsars released during the first and second explosion, as well as the frequency of 
surviving binaries depends sensitively on the magnitude distribution of 
natal SN kicks. A binary system  with a given separation is more likely to survive a kick of a certain magnitude during the first SN explosion, when the secondary is still a massive star. The second SN can then release up to two NSs (one young and one midly recycled). In this case one would expect a relatively large number of isolated mild MSPs, which appears to disagree with observations ($f_{\rm mild}^{\rm obs}\simeq 85\%$, see Section~\ref{sec:4}). 
This suggests that most young pulsars originate from the first SN  in a wide binary, or the released secondary star.
On the other hand, compact orbits, stripping of the secondary star prior to its core collapse and electron-capture SNe lead to increased  binary survival rates and therefore, they can help explain the properties of the observed binary pulsar population \citep{podsiadlowski2004a,tauris2013,tauris2015,Tauris:2017omb}.   
Nevertheless our results appear to be in broad agreement with kick amplitudes of few 100\,km\,s$^{-1}$ for core-collapse SNe \citep{schneider2020a}, while extremely small-amplitude kicks are disfavoured given the  small number of pulsars in wide binaries \citep[see also][]{Igoshev:2019cwq}. Detailed population-synthesis studies and comparison with all available observational benchmarks will be required to reach more  quantitative conclusions. 

It is evident that this study will benefit significantly by future iterations of the \gaia\ catalogue that will provide improved astrometry for a larger number of stars. Next-generation  surveys such as the \emph{Vera C. Rubin Large Synoptic Survey Telescope} \citep{Kessler:2019qge} will be even more important in identifying dimmer and less massive pulsar companions (Figure~\ref{sec:3}) as well as optical emission from young pulsars similar to the Crab. Close astrometric pairs --- even those that are not true binary companions ---  can also be promising targets for pulsar scintillation studies \citep{simard2018,main2018,main2020}. Finally, an extension of the method presented here combined with improved astrometry will also allow to identify potential future micro-lensing events \citep{Dai:2015jua} and stars ejected during SNe.  

Despite the current importance of optical astrometry in constraining the frequency of binary pulsars, our simulation suggest that  
pulsar timing remains the  most sensitive probe of multiplicity. Until recently, regular monitoring of all known pulsars was impractical due to limited 
telescope availability and resources. The advent of large-scale monitoring  
projects such as the CHIME/pulsar project \citep{ng2018,ng2020} and its southern counterpart HIRAX \citep{newburgh2016}, MeerTime  \citep{bailes2016,bailes2020,Johnston:2020qxo} and eventually the Square Kilometre Array \citep{tauris2015a,antoniadis2015} will therefore lead to significantly more stringent constraints on the young pulsar multiplicity fraction in the near future.

\section*{Data availability}
The python/jupyter notebook  to reproduce the results of this article is available at \url{https://dx.doi.org/10.5281/zenodo.4075042}.

\section*{Acknowledgements} 
Support was provided by the Hellenic Foundation for
Research and Innovation (HFRI) and the Stavros Niarchos Foundation (SNF) under grant agreement No.\,72-1/11.8.2020. I am grateful to the referee for their constructive report. I am also grateful to Philipp Podsiadlowski, Paulo Freire, Michael Kramer and Vivek Venkatraman Krishnan for discussions. This work relies on data from the
European Space Agency (ESA) mission \gaia\
(\url{https://www.cosmos.esa.int/gaia}), processed by the
\gaia\ Data Processing and Analysis Consortium (DPAC,
\url{https://www.cosmos.esa.int/web/gaia/dpac/consortium}). 
This research made use of NumPy \citep{2020NumPy-Array}, Matplotlib \citep{Matplotlib} and  Astropy (\url{http://www.astropy.org}), a community-developed core Python package for Astronomy \citep{Robitaille:2013mpa,Price-Whelan:2018hus} and of NASA's Astrophysics Data System Bibliographic Services.

\begin{landscape}
\begin{table}
\caption{Catalogued and derived properties of confirmed and candidate cross-matches between the \gaia\ DR2 and the ATNF pulsar catalogues. A complete catalogue of all \gaia\ sources within 20\arcsec\ of ATNF pulsars, including more columns is available online (see Appendix).}
\small{
\begin{tabular}{lcccccccccccccccl}
\hline
\multicolumn{15}{c}{ Known Binaries} \\
\hline 
Name & $\theta$ & $\sigma_{\rm pos}$ & $\hat{\pi}_{\rm GAIA}$ &  $\hat{\pi}_{\rm r}$ & $\mu_{\alpha}$ & $\mu_{\delta}$ & $\mu_{\alpha,r}$ & $\mu_{\delta,r}$ & $P_s$ & P$_{\rm b}$ & DM & m$_{\rm g}$ & P$_{\rm assoc.}$ & d & M$_{\rm g}$\\
& \multicolumn{2}{c}{ arcsec } & \multicolumn{2}{c}{ mas } & \multicolumn{4}{c}{ mas\,yr$^{-1}$ } & s & d & pc\,cm$^{-3}$ & & & kpc & \\

\hline
J0045$-$7319 & 0.5149 & 0.2689 & 0.04(6) &- & 0.30(11)& $-0.9(1)$ & - & - & 0.926 & 51.169 & 105.4 & 16.22 & 0.90 &$62.1^{+1.9}_{-1.9}$ & $-5.30^{+0.06}_{-0.07}$ \\
J0337+1715 & 0.0064 & 0.2501 & 0.73(25) &- & 4.8(5) & $-4.4(4)$ & - & - & 0.003 & 1.629 & 21.3 & 18.08 & 0.99  & $1.49^{+0.66}_{-0.47}$ & $6.88^{+0.79}_{-0.82}$\\
J0348+0432 & 0.0018 & 0.2519 & $-1.9(12)$ &- & 3.1(20) & $-0.1(1.4)$ & 4.04(16) & 3.5(6) & 0.039 & 0.102 & 40.5 & 20.64 & 0.99 & $2.49^{+1.17}_{-1.01}$ & $7.86^{+0.84}_{-1.13}$ \\
J0437$-$4715 & 0.3286 & 0.2506 & 8.32(68) & 6.37(9) & 122.8(12) & $-71.2(1.7)$ & 121.438(2) & $-71.475(2)$ & 0.006 & 5.741 & 2.6 & 20.41 & 0.99 & $0.157^{+0.003}_{-0.004}$ & $14.43^{+0.04}_{-0.06}$ \\
J1012+5307 & 0.0165 & 0.2501 & 1.32(41) & 0.71(17) & 2.97(52) & $-26.94(63)$ & 2.609(8) & $-25.482(11)$ & 0.005 & 0.605 & 9.0 & 19.63 & 0.99 &  $1.73^{+1.62}_{-0.58}$ & $8.40^{+1.43}_{-0.89}$ \\
J1023+0038 & 0.0042 & 0.2500 & 0.73(14) & 0.731(22) & 4.75(14) & $-17.35(14)$ & 4.76(3) & $-17.34(4)$ & 0.002 & 0.198 & 14.3 & 16.27 & 0.99 &  $1.371^{+0.072}_{-0.065}$ & $5.51^{+0.11}_{-0.10}$\\
J1024$-$0719 & 0.0765 & 0.2503 & 0.53(43) & 0.8(3) & $-35.52(64)$ & $-47.93(66)$ & - & - & - &0.005 & 6.5 & 19.18 & 0.99 & $3.4^{+4.6}_{-2.0}$ & $6.41^{+1.86}_{-1.99}$ \\
J1048+2339 & 0.0103 & 0.2511 & 0.96(80) &- & $-16.3(10)$ & $-11.70(13)$ & $-18.7$ & $-9.4$ & 0.005 & 0.251 & 16.7 & 19.65 & 0.99 & $2.1^{+1.2}_{-1.0}$ & $7.99^{+0.97}_{-1.45}$ \\
J1227$-$4853 & 0.0618 & 0.2500 & 0.62(17) &- & $-18.7(2)$ & 7.4(1) & - & - & 0.002 & 0.288 & 43.4 & 18.08 & 0.99 & $2.0^{+2.2}_{-0.7}$ & $6.41^{+1.69}_{-1.04}$ \\
J1311$-$3430$^{\ast}$ & 0.0416 & 2.89 & - &- &- &- &- & -& 0.002 & 0.065 & 37.84 & 20.52 & - &  $2.43^{+0.48}_{-0.48}$ & $8.54^{+0.41}_{-0.50}$\\
B1259$-$63 & 0.0590 & 0.2500 & 0.42(3) &- & $-6.986(43)$ &$-0.416(44)$ & $-6.6(18)$ & $-4.4(14)$ & 0.048 & 1236.725 & 146.7 & 9.63 & 0.96 & $2.42^{+0.33}_{-0.26}$ & $-7.05^{+0.94}_{-0.77}$ \\
J1417$-$4402 & 0.3818 & 0.2805 & 0.22(7) & - & $-4.7(1)$ & $-5.1(1)$ & - & - & 0.003 & 5.374 & 55.0 & 15.79 & 0.99 &  $4.6^{+3.1}_{-1.5}$ & $2.25^{+1.26}_{-0.93}$\\
J1431$-$4715 & 0.0485 & 0.2500 & 0.64(16) &- & $-12.01(33)$ & $-14.50(26)$ & $-7(3)$ & $-8(4)$ & 0.002 & 0.450 & 59.4 & 17.75 & 0.99 & $2.0^{+2.1}_{-0.7}$ & $6.13^{+1.68}_{-1.03}$\\
J1723$-$2837 & 0.3122 & 0.2500 & 1.077(54) &- & $-11.71(8)$ & $-23.99(6)$ & - & - & 0.002 & 0.615 & 19.7 & 15.55 & 0.93 & $0.94^{+0.086}_{-0.073}$ & $4.19^{+0.20}_{-0.18}$\\
J1810+1744 & 0.0689 & 0.2519 & 1.04(70) &- & 6.4(18) & $-7.2(21)$ & - & - & 0.002 & 0.150 & 39.7 & 20.08 & 0.99 & $2.3^{+1.4}_{-1.2}$ & $7.76^{+1.02}_{-1.70}$ \\
J1816+4510 & 0.0016 & 0.2501 & 0.22(15) & -& $-0.16(30)$ & $-4.41(33)$ & 5.3(8) & $-3(1)$ & 0.003 & 0.361 & 38.9 & 18.22 & 0.99 &  $4.2^{+2.2}_{-1.6}$ & $5.03^{+0.92}_{-1.056}$ \\
J1957+2516 & 0.0066 & 0.2512 & 0.69(86) &- & $-5.7(11)$ & $-8.8(15)$ & - & - & 0.004 & 0.238 & 44.1 & 20.30 & 0.96 &  $2.6^{+1.5}_{-1.3}$ & $5.47^{+1.96}_{-2.59}$ \\
B1957+20 & 0.1016 & 0.2509 & $-0.6(16)$ &- & $-19.1(12)$ & $-23.5(19)$ & $-16.0(5)$ & $-25.8(6)$ & 0.002 & 0.382 & 29.1 & 20.30 & 0.96 & $1.9^{+1.0}_{-0.9}$ & $7.88^{+1.06}_{-1.56}$\\
J2032+4127 & 0.0480 & 0.2500 & 0.69(3) &- & $-2.991(48)$ & $-0.742(55)$ & - & - & 0.143 & 16835.000 & 114.7 & 11.36 & 0.99 &  $1.45^{+0.13}_{-0.11}$ & $-3.19^{+1.51}_{-1.40}$\\
J2129$-$0429$^{\ast}$ & 1.388 & 0.2500 &0.424(88) & - & 12.38(15)& 10.18(15) &- &- & 0.007 & 0.635 & 16.9  & 16.83  & - &$2.60^{+1.48}_{-0.71}$ & $4.69^{+0.98}_{-0.70}$\\
J2215+5135 & 0.0887 & 0.2502 & 0.27(37) &- & 0.3(5) & 1.9(6) & 170(22) & 82(24) & 0.003 & 0.173 & 69.2 & 19.24 & 0.98 & $2.9^{+1.5}_{-1.2}$ & $6.46^{+1.17}_{-1.35}$ \\
J2339$-$0533 & 0.1247 & 0.2505 & 0.75(26) &- & 4.1(5) & $-10.3(3)$ & 11(4) & $-29(10)$ & 0.003 & 0.193 & - & 18.97 & 0.99 &  $1.35^{+0.55}_{-0.40}$ & $8.27^{+0.73}_{-0.76}$\\
\hline
\multicolumn{15}{c}{Single Pulsars} \\
\hline
B0531+21 & 0.0268 & 0.2500 & 0.27(12) &- & $-11.82(22)$ & 2.65(17) & $-14.7(8)$ & 2.0(8) & 0.033 &- & 56.8 & 16.43 & 0.99 & $4.0^{+3.6}_{-1.6}$ & $2.14^{+1.64}_{-1.16}$\\
B0540$-$69 & 0.0163 & 0.2503 & $-0.53(14)$ & -& 2.8(28) & 2.1(32) & - & - & 0.051 &- & 146.5 & 20.31 & 0.93 & $62.1^{+1.89}_{-1.89}$ & $-6.30^{+0.06}_{-0.07}$ \\
\hline
\multicolumn{15}{c}{Small angular separations  but likely not associated} \\
\hline
B1953+29 & 0.1401 & 0.2501 & 0.60(19) &- & $-3.1(3)$ & $-7.96(32)$ & - & - & 0.006 & 117.349 & 104.5 & 18.71 & 0.97 & $2.3^{+2.9}_{-1.0}$ & $4.54^{+2.70}_{-1.91}$\\
J1435$-$6100 & 0.3993 & 0.2500 & $-0.26(35)$ & -& $-5.28(33)$ & $-2.34(45)$ & - & - & 0.009 & 1.355 & 113.7 & 18.94 & 0.95 & $3.3^{+1.5}_{-1.2}$ & $-2.70^{+4.92}_{-4.45}$\\
\hline
\multicolumn{15}{c}{ Candidates } \\
\hline
J0534$-$6703 & 0.5130 & 0.2872 & 0.14(18) & - & 1.7(5) & 0.3(5) & - & - & 1.818 &- & 94.7 & 18.90 & 0.94 & $49.97^{+1.13}_{-1.13}$ & $-0.12^{+0.05}_{-0.05}$ \\
J1624$-$4411 & 0.3068 & 0.2514 & 1.14(59) &- & $-2.6(11)$ & $-8.6(8)$ & - & -  & 0.233 & - & 139.4 & 19.93 & 0.94 & $2.7^{+4.6}_{-1.8}$ & $6.32^{+4.63}_{-3.38}$ \\
J1638$-$4608 & 0.1407 & 0.2553 &- &- &- &- &- & -  & 0.278 &- & 423.1 & 19.38 & 0.93 &- &- \\
J1838$-$0549 & 0.5015 & 0.2536 & 0.65(19) &- & 1.55(26) & 1.62(23) & - & - & 0.235 &- & 276.6 & 17.71 & 0.98 & $2.0^{+2.5}_{-0.8}$ & $1.73^{+2.84}_{-1.60}$\\
J1903$-$0258 & 0.4002 & 0.2505 & 0.95(30) &- & $-1.7(6)$ & 3.40(46) & - & - & 0.301 &- & 113.0 & 18.93 & 0.96 & $1.7^{+3.2}_{-0.8}$ & $7.55^{+5.49}_{-1.53}$ \\
J1958+2846 & 0.3912 & 0.2537 & 0.59(34) &- & $-4.0(5)$ & $-6.1(6)$ & - & - & 0.290 &- &- & 19.36 & 0.98 & $2.0^{+1.0}_{-0.8}$ & $5.34^{+1.45}_{-1.71}$\\
J2027+4557 & 0.2904 & 0.2500 & 0.52(3)  &- & 2.12(7) & $-2.06(6)$ & - & - & 1.100 &- & 229.6 & 15.73 & 0.98 & $1.96^{+0.25}_{-0.20}$ & $1.60^{+0.32}_{-0.56}$\\
J1852+0040 & 0.5067 & 0.2571 & 2.1(12) &- & -0.9(17) & -4.9 & - & - & 0.105 &- &- & 20.23 & 0.98 & $2.93^{+4.83}_{-2.20}$ & $3.23^{+4.78}_{-4.14}$ \\
\hline
\end{tabular}\label{tab:1}
($^\ast$) Sources identified only after visual inspection
}
\end{table}
\end{landscape}

\appendix

\section{Description of columns in online catalogue}

The full catalogue of sources withing 20\arcsec\ of ATNF pulsars is available on VizieR (\url{http://cdsarc.u-strasbg.fr/viz-bin/}). The table below provides a summary of the information available in the catalogue.

\begin{table*}
\caption{Catalogue Format}
\begin{tabular}{ccc}
\hline 
Name & Format  & Description \\
\hline
\texttt{angDist} & float64 &  angular Separation from the pulsar  (in degrees)  \\
\texttt{id} & int64 &  pulsar identification number  \\
\texttt{name} & <U12 &  pulsar name \\
\texttt{survey} & <U48 & survey that discovered the pulsar as defined in psrcat \\
\texttt{BinComp} & <U6 & companion type as defined in psrcat \\
\texttt{Type} & int64 & pulsar type as defined in psrcat \\
\texttt{raj} & float64 &  pulsar right ascension at \texttt{PosEpoch}  (in degrees) \\
\texttt{rajerr} & float64 &  error in pulsar right ascension at \texttt{PosEpoch}  (in degrees)  \\
\texttt{decj} & float64 &   pulsar right ascension at \texttt{PosEpoch}  (in degrees)\\
\texttt{decjerr} & float64 &  error in pulsar right ascension at \texttt{PosEpoch}  (in degrees) \\
\texttt{pmra$\_$radio} & float64 &  radio timing proper motion in RA (mas/yr) \\
\texttt{pmraerr} & float64 &  error of radio timing proper motion in RA (mas/yr) \\
\texttt{pmdec\_radio} & float64 &  radio timing proper motion in DEC (mas/yr) \\
\texttt{pmdecerr} & float64 & error of radio timing proper motion in DEC (mas/yr)  \\
\texttt{px} & float64 & radio timing parallax (mas) \\
\texttt{pxerr} & float64 & error of radio timing parallax (mas) \\
\texttt{posepoch} & float64 & position epoch of the radio timing solution \\
\texttt{p0} & float64 & spin period (s) \\
\texttt{p0err} & float64 &  error of spin period (s) \\
\texttt{p1} & float64 & spin-period derivative \\
\texttt{p1err} & float64 & error of spin-period derivative \\
\texttt{dm} & float64 &  dispersion measure (pc/cm$^3$)\\
\texttt{dmerr} & float64 & error of  dispersion measure (pc/cm$^3$) \\
\texttt{dist} & float64 & DM distance estimate (see text) \\
\texttt{pb} & float64 & orbital period (days)  \\
\texttt{pberr} & float64 & error of orbital period (days) \\
\texttt{raj\_now} & float64 &  RA of source at \texttt{posepoch} \\
\texttt{decj\_now} & float64 & DEC of source at \texttt{posepoch}  \\
\texttt{ra\_epoch2000} & float64 &   RA of source at \texttt{J2000}\\
\texttt{dec\_epoch2000} & float64 &   DEC of source at \texttt{J2000} \\
\texttt{source\_id} & int64 & GAIA source ID \\
\texttt{ra} & float64 &  RA of source at \texttt{J2015.5} \\
\texttt{ra\_error} & float64 & RA uncertainty of source at \texttt{J2000}  \\
\texttt{dec} & float64 &  DEC of source at \texttt{J2000} \\
\texttt{dec\_error} & float64 &  DEC uncertainty of source at \texttt{J2000} \\
\texttt{parallax} & float64 &  \gaia\ parallax (mas) \\
\texttt{parallax$\_$error} & float64 & \gaia\ parallax error (mas) \\
\texttt{pmra} & float64 & \gaia\ proper motion in RA (mas/yr) \\
\texttt{pmra\_error} & float64 & error of \gaia\ proper motion in RA (mas/yr) \\
\texttt{pmdec} & float64 & \gaia\ proper motion in DEC (mas/yr) \\
\texttt{pmdec\_error} & float64 &  error of \gaia\ proper motion in DEC (mas/yr) \\
\texttt{phot\_g\_mean\_mag} & float64 &  \gaia\ mean magnitude  \\
\texttt{phot\_bp\_mean\_mag} & float64 & \gaia\ mean magnitude in BP band  \\
\texttt{phot\_rp\_mean\_mag} & float64 &  \gaia\ mean magnitude in RP band  \\
\texttt{bp\_rp} & float64 &  \gaia\ BP-RP mean color   \\
\texttt{poserror} & float64 & position error of the source at \texttt{posepoch} (arcsec) \\
\texttt{poserror\_sys} & float64 &  position error of the source including systematics at \texttt{posepoch} (arcsec) \\
\texttt{angDist} & float64 & angular separation from the nearest pulsar (arcsec) \\
\texttt{TrueAngDist} & float64 &  angular separation from the nearest pulsar at \texttt{posepoch} (arcsec) \\
\texttt{Pass} & float64 & probability of Association (1 - Probability of chance coincidence) \\
\texttt{g\_abs}$^{\ast}$ & float64 & Absolute G magnitude corrected for extinction\\
\texttt{b\_r\_abs}$^{\ast}$ & float64 &  $B-R$ color corrected for extinction\\
\texttt{distance}$^{\ast}$ & float64 & Median distance (kpc) \\
\texttt{distance\_min}$^{\ast}$ & float64 & 16\% lower bound on distance (kpc) \\
\texttt{distance\_max}$^{\ast}$ & float64 & 68\% upper bound on distance (kpc) \\
\texttt{dust} & float64 & median E(B-V) extinction \\
\texttt{dust\_min} & float64 & 16\% lower bound on extinction \\
\texttt{dust\_max} & float64 & 68\% upper bound on extinction \\
\hline
\end{tabular} \\
$^{\ast}$ Calculated only for the sources in Table\,\ref{tab:1}
\end{table*}

\bibliographystyle{mnras}
\bibliography{gaiapsr} 






\bsp	
\label{lastpage}
\end{document}